\documentclass{aa}  
\usepackage{graphicx}
\usepackage{subfigure}

\usepackage{txfonts}
\begin{document}

   \title{Hot methanol from the inner region of the HH\,212 protostellar system}

    \author{S. Leurini\inst{1}
      \and C. Codella\inst{2}
      \and S. Cabrit\inst{3}
      \and F. Gueth\inst{4}
      \and A. Giannetti\inst{1}
      \and F. Bacciotti\inst{2}
      \and R. Bachiller\inst{5}
      \and C. Ceccarelli\inst{6}
      \and A. Gusdorf\inst{7}
      \and B. Lefloch\inst{6}
      \and L. Podio\inst{2}
      \and M. Tafalla\inst{5}
          }

   \institute{Max-Planck-Institut f\"ur Radioastronomie, Auf dem H\"ugel 69, D-53121, Bonn, Germany\\
   \email{sleurini@mpifr.de}
   \and INAF, Osservatorio Astrofisico di Arcetri, Largo E. Fermi 5, 50125 Firenze, Italy
 \and LERMA, Observatoire de Paris, PSL Research University, CNRS, Sorbonne Universit\'es, UPMC Univ. Paris 06, \'Ecole Normale Sup\'erieure, F-75014, Paris, France
   \and IRAM, 300 rue de la Piscine, 38406 Saint Martin d’H\`eres, France
   \and IGN, Observatorio Astronómico Nacional, Alfonso XII 3, 28014 Madrid, Spain
      \and Univ. Grenoble Alpes, CNRS, IPAG, F-38000, Grenoble, France      
   \and LERMA, D\'epartement de physique de l’ENS, ENS, Observatoire de Paris, PSL Research University, Universit\'e Cergy-Pontoise, Universit\'e Paris Seine, Sorbonne Universit\'es, UPMC Univ Paris 06, CNRS, Paris France}
     
   \date{}

   \abstract{The mechanisms leading to the formation of  disks around  young stellar objects (YSOs) and to the launching of the associated jets  are crucial to the understanding of the earliest stages of star and planet formation.  HH\,212 is a privileged laboratory to study a pristine jet-disk system. Therefore we investigate the innermost region ($<100$\,AU) around the HH\,212-MM1 protostar through ALMA band\,7 observations of methanol.
     The 8\,GHz bandwidth spectrum towards the peak of the continuum emission of the HH\,212 system reveals at least 19 transitions of methanol. Several of these lines (among which several vibrationally excited lines in the $\varv_{\rm t}=1,2$ states) have upper energies above 500\, K. They originate from a compact ($<135$\, AU in diameter), hot ($\sim 295$\,K) region elongated along the direction of the SiO jet.  We performed a fit in the $uv$ plane of various velocity channels of the strongest high-excitation lines. The blue- and red-shifted velocity centroids are shifted roughly symmetrically on either side of the jet axis, indicating that the line-of-sight velocity beyond 0.7\,km\,s$^{-1}$ from systemic is dominated by rotational motions. The  velocity increases moving away from the protostar further indicating that the emission of methanol is not associated with a Keplerian disk or rotating-infalling cavity, and it is more likely associated with outflowing gas. We speculate that  CH$_3$OH traces a disk wind gas accelerated
at the base. The launching region would be at a radius of  a few astronomical units from the YSO.}


   \maketitle
\section{Introduction}
The Class 0 protostellar stages are clearly associated with
mass loss and in particular with fast collimated jets, usually
observed as extremely high-velocity structures in CO and SiO
transitions.  In contrast,
the evidence of disks in these very early phases of evolution is much
less clear. Magnetic fields play a fundamental role in regulating the
formation of  young stellar objects (YSO) and of the disk as they
are believed to remove the excess  angular momentum from the
infalling material allowing accretion onto the central
object. However, this ``magnetic braking'' is so efficient that
Keplerian disks may be initially suppressed beyond 10\,AU
\citep{2007Ap&SS.311...75P}. The expected small sizes of disks in Class 0 phases and the fact that in these still deeply embedded objects the emission of the surrounding  envelope  is likely 
entangled with that of the disk  make the detection of Keplerian disks in Class 0 YSOs very challenging \citep[e.g.][]{2009ApJ...699.1584L,2012Natur.492...83T} even in the ALMA era \citep[e.g.][]{2013A&A...560A.103M,2014Natur.507...78S,2014ApJ...796..131O}.

HH\,212 is a strikingly bright and symmetric bipolar jet from a Class 0 source at a distance of 450\,pc extensively observed with IRAM PdBI, SMA, and  ALMA \citep{2006ApJ...639..292L,2007ApJ...659..499L,2007A&A...462L..53C,2007A&A...468L..29C,2008ApJ...685.1026L,2012A&A...548L...2C,2014A&A...568L...5C,2015ApJ...805..186L,2015A&A...581A..85P}. 
A disk was observed with ALMA  
towards the HH\,212--MM1 protostar  
in HCO$^+$, C$^{17}$O, and SO emission with velocity gradients 
along the equatorial plane consistent with a rotating 
disk of $\simeq 0\farcs2$ = 90 AU in radius 
around a $\simeq 0.2-0.3 M_{\rm \odot}$ source.
Therefore, the HH\,212 region is, to our knowledge, 
the only example of a protostellar region with a bright 
bipolar jet and a compact rotating disk, and is thus a privileged laboratory to study
a pristine jet-disk system.
The asymmetric line profile of one high-excitation ($\sim$300 K) HDO transition recently observed towards HH\,212--MM1 \citep{2016A&A...586L...3C}  suggests the possible occurrence of hot and expanding gas associated with a disk wind, calling for further multiline observations.

In this Letter, we
further exploit ALMA Band 7 data \citep[from][]{2014A&A...568L...5C} to investigate the inner 100 AU of
the HH\,212 system through a survey of  methanol (CH$_3$OH) high-energy lines (up to $E_{\rm u}$ = 747 K).  High-excitation lines of CH$_3$OH  probe the  innermost gas around the protostar \citep[e.g.][]{2007A&A...475..925L,2014A&A...563L...1M} and can be  used as a selective tracer of the  kinematics of the region.

%--------------------------------------------------------------------

\section{Observations}
         The data presented  are part of the observations discussed by \citet{2014A&A...568L...5C}, \citet{2015A&A...581A..85P}, and \citet{2016A&A...586L...3C}. We refer to these papers for further details and give here a short summary of the observations. 
HH\,212  was observed in Band 7
during the cycle 0 phase of ALMA.
The data cover the frequencies 333.7--337.4\,GHz and 345.6--349.3\,GHz
with a spectral resolution of  488\,kHz 
(corresponding to 0.42--0.44\,km\,s$^{-1}$).
The continuum-subtracted images have a typical
clean-beam FWHM of $0\farcs6\times0\farcs5$
(PA $\simeq$ 40$\degr$), and an rms noise level of
3\,mJy\,beam$^{-1}$ in the 488 kHz channel.
Positions are given with respect to the  continuum peak
($\alpha({\rm J2000})$ = 05$^h$ 43$^m$ 51$\fs$41,
$\delta({\rm J2000})$ = --01$\degr$ 02$\arcmin$ 53$\farcs$17,
\citealt{2014ApJ...786..114L}).

\section{Results}
\begin{figure}[]
\centering
\includegraphics[width=8cm]{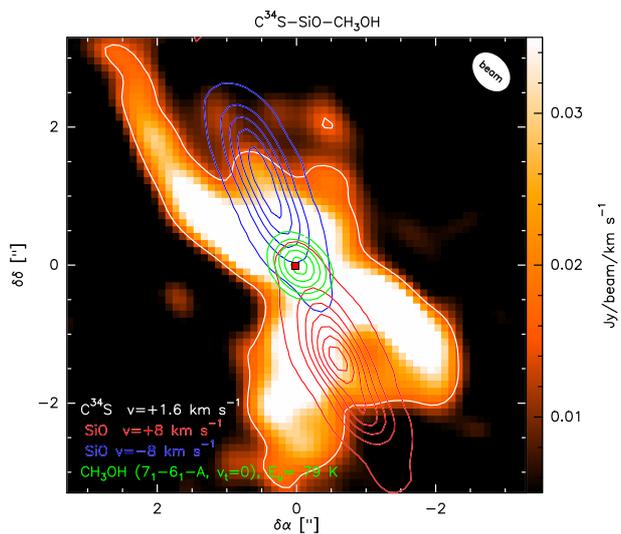}
\caption{The  HH\,212  protostellar  system  as  observed by ALMA \citep{2014A&A...568L...5C}. The colour scale  represents the C$^{34}$S (7--6) emission close to the systemic velocity (the white contour is the  5\,$\sigma$ level).   Blue and red contours  show  the  blue-
  and red-shifted  SiO(8–7)  jet  at $\pm$8 \,km\,s$^{-1}$ from the systemic velocity. Green contours show the integrated emission of the $  7_1 - 6_1$-$A$,\,$\varv_t=0$ CH$_3$OH line (from 10\,$\sigma$, 90\,mJy\,beam$^{-1}$\,km$^{-1}$\,s, in steps of 10\,$\sigma$).
The red square marks  the peak of the continuum emission \citep{2014ApJ...786..114L}. The filled ellipse shows the synthesised  beam.}\label{fig1}
\end{figure}

\begin{figure}[]
\includegraphics[width=0.9\columnwidth]{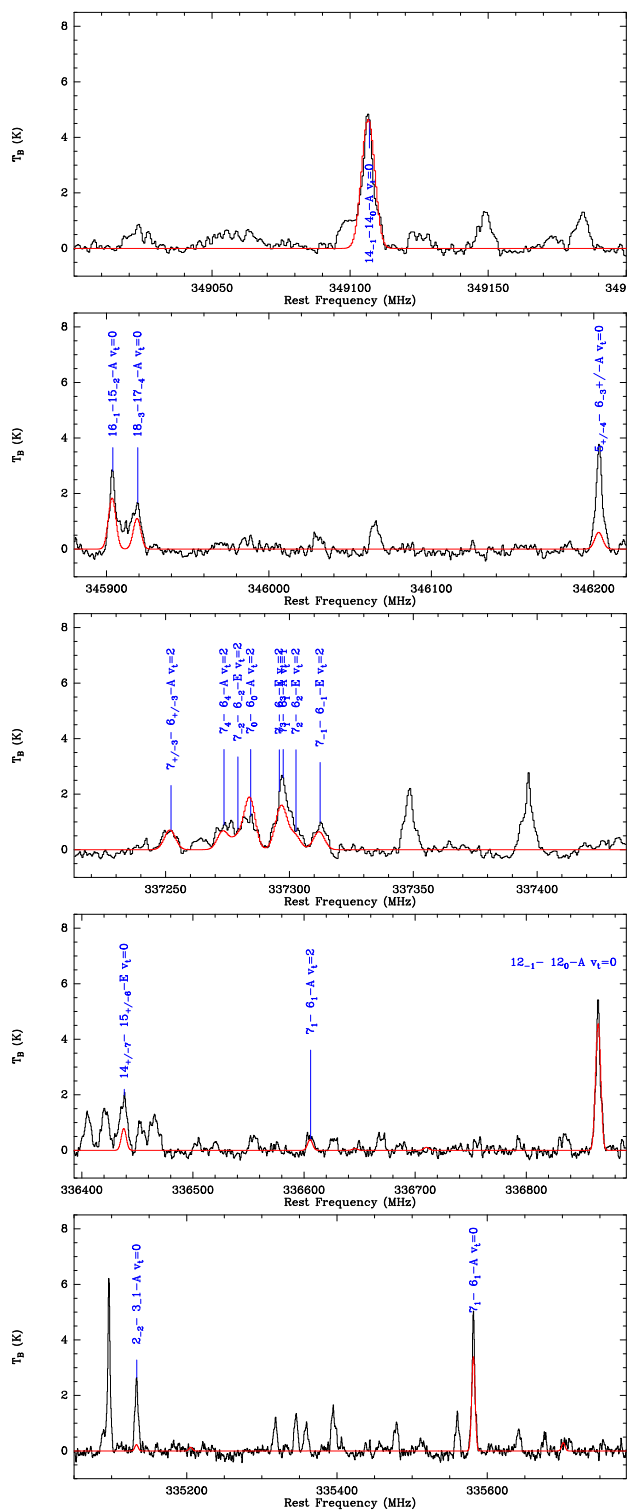}
\caption{Zoom-in of the different regions of the spectrum  shown in Fig.\,\ref{figa1}. The best-fit LTE CH$_3$OH synthetic spectrum is displayed in red and
overlaid  on  the  observed  spectrum   shown  in  black. 
The rest frequency of the fitted CH$_3$OH transitions listed in Table\,\ref{tab_id} are labelled in blue.}\label{fig2}
\end{figure}

The HH\,212 protostellar system is shown in Fig.\,\ref{fig1}. 
We identified 19 lines of CH$_3$OH in the 8\,GHz
bandwidth spectrum towards  the  0.9\,mm continuum peak (see Figs. \ref{fig2} and\,\ref{figa1} for the full spectrum)
 using the Weeds package
\citep{2011A&A...526A..47M} and the spectroscopic parameters from the
Jet Propulsion Laboratory (JPL) molecular database
\citep{pickett_JMolSpectrosc_60_883_1998}.  Typical line-widths are 4--5\,km\,s$^{-1}$.
The identified lines are in Table\,\ref{tab_id} and  they  belong to 
the first three vibrational states: nine lines in the $\varv_t=0$ level, one in $\varv_t=1$, and
nine in $\varv_t=2$, respectively. 
To our knowledge,  $\varv_t=1$ CH$_3$OH lines around low-mass YSOs were previously reported only around the Class 0 protostar NGC\,1333--IRAS2A \citep{2014A&A...563L...1M,2014A&A...563L...2M}. 
The lines cover a wide excitation range
($E_{\rm u}\sim$79\,K to 747\,K) and
they  sample  the high-excitation regime well (eight transitions
have $E_{\rm u}>500$\,K ). These data
therefore offer a unique opportunity to investigate the nature of
the hot gas surrounding the protostar.

\subsection{High-excitation CH$_3$OH emission}

Figure\,\ref{fig1} shows the integrated intensity emission of the $7_1-6_1$-$A$,\,$\varv_t=0$
line;  integrated maps of
  other lines not affected by severe overlapping with other spectral features are shown in Fig.\,\ref{fig7}. The emission is clearly compact and indeed traces 
the inner region close to the protostar.
We averaged the visibilities  over the range $\pm2$\,km\,s$^{-1}$ from the peak velocity of the $7_1-6_1$-$A$,\,$\varv_t=0$, $ 12_{-1}  - 12_0$-$A$,\,$\varv_t=0$,
and $14_{-1} - 14_0$-$A$,\,$\varv_t=0$ lines to determine the size of the emitting region assuming an elliptical Gaussian distribution in space  using the  GILDAS $uv$-fit task.
These lines are the 
strongest and most isolated in our dataset and are therefore less affected by blending with other transitions.  The $\pm2$\,km\,s$^{-1}$ velocity range   minimises blending with a  spectral feature at red-shifted velocities  for the $14_{-1} - 14_0$-$A$,\,$\varv_t=0$ line (Fig.\,\ref{fig2}).
The fit   gives a typical size of  
0$\farcs2- 0\farcs$3 (90\,AU--135\,AU in diameter, see Table\,\ref{tab_uvfit} and Fig.\,\ref{fig4})   in good agreement 
with the estimate from CH$_3$CHO  of  \citet{2016A&A...586L...3C} with the
same ALMA dataset. The position angles of the Gaussian fit are consistent  within the errors with the position angle of the SiO jet ($22^\circ$). This suggests that CH$_3$OH is associated with the jet/outflow system. Interestingly, the size of the emitting region seems to decrease going towards higher energies as found by \citet{2014A&A...563L...2M} in NGC\,1333--IRAS2A for  complex molecules. Since the  analysed lines are the strongest of the dataset and have similar peak intensities, we  believe that these results are not biased by different signal-to-noise levels.

\begin{table}
\caption{Results of the $uv$ fit of the averaged\tablefootmark{a} visibilities of  CH$_3$OH lines\tablefootmark{b}.}   
\label{tab_uvfit}
\centering
\begin{tabular}{l r r c c}  
\hline\hline                 
Transition &$\Delta\alpha\tablefootmark{c}$ &$\Delta\delta\tablefootmark{c}$&FWHM\tablefootmark{d}&P.A.\tablefootmark{e}\\
           &    (mas)            &    (mas)     &       & \\
\hline                       
$ 7_1    - 6_1$-$A$,\,$\varv_t=0$   & -12 &  +0 &$0\farcs26\times0\farcs19$&$36^\circ$\\
$ 12_{-1}  - 12_0$-$A$,\,$\varv_t=0$& -19 &  +0 &$0\farcs23\times0\farcs16$&$32^\circ$\\
$ 14_{-1} - 14_0$-$A$,\,$\varv_t=0$ & -7 & +12 &$0\farcs20\times0\farcs14$&$16^\circ$\\
\hline 
\end{tabular}
\tablefoot{\tablefoottext{a}{Averaged over [-2,+2]\,km\,s$^{-1}$ from the peak.}
  \tablefoottext{b}{The resulting error on the centroid position is  a function of the channel signal-to-noise ratio and atmospheric seeing, and is typically much smaller than the beam size.}
  \tablefoottext{c}{Offsets are with respect to the peak of the continuum emission. Uncertainties are $\sim$ 4 mas.}
  \tablefoottext{d}{Uncertainties  are $\sim$ 30 mas.}
  \tablefoottext{e}{Uncertainties  are  $\sim10^\circ-18^\circ$.}
}
\end{table}

To extract physical parameters from the data, a simultaneous fit was performed on the full 
spectrum. Weeds  generates synthetic
spectra of a given molecule assuming local thermodynamic equilibrium (LTE)
 over the full observed bandwidth.
Since Weeds lacks  an automatic optimisation
algorithm  we used MCWeeds (Giannetti et al. in prep.), an external interface between
Weeds and PyMC \citep{Patil+10_jstatsoft35_1},  to implement Bayesian statistical models
and fitting algorithms. We fit all CH$_3$OH transitions   using the Markov chain Monte Carlo method. We used 100\,000 iterations following a burn-in period of 20\,000 with a thinning factor of 50. 
The source
size was kept as a fixed parameter (0$\farcs$2, Table\,\ref{tab_uvfit}). 
Our analysis indicates that (i) CH$_3$OH is optically
thin ($\tau\le0.4$), (ii) the best LTE fit temperature is well constrained to $\simeq$ 295\,K, and
(iii) the total column density at this temperature is 
$N_{\rm CH_3OH} \simeq 3\times10^{17}$ cm$^{-2}$. The best fit results and the 95\%
highest probability density (HPD) interval ranges are given in Table\,\ref{tab_lte}; the synthetic spectra corresponding to the best fit are shown in Fig.\,\ref{fig2}. The fit reproduces all  lines except the $  2_{-2} - 3_{-1}$-$A$,\,$\varv_t=0$ and $  5_{\pm4} - 6_{\pm3}$-$A$,\,$\varv_t=0$ transitions which are  underestimated by the model. These lines are among the lowest in energy  in the dataset and likely trace a lower temperature regime than the other transitions.

 The bolometric luminosity of HH\,212 is 
 $\sim14$\,$L_\odot$ \citep{1992A&A...265..726Z}.
Using Eq.\,1 of \citet{2000A&A...355.1129C}, the dust temperature would  be about 250\,K at a
radius of 10\,AU, a factor of 5 lower than the size inferred from
the present  data. This suggests that the gas is thermally decoupled from dust, as was found for   the inner regions of low-mass protostellar envelopes \citep[e.g.,][]{1996ApJ...471..400C,2009A&A...506.1229C} and/or that the CH$_3$OH excitation is enhanced by absorption of the IR radiation field. 
Indeed, 
\citet{2007A&A...466..215L} showed
that CH$_3$OH $\varv_t=1$  lines   trace  the IR radiation field of the protostar (see their Fig.\,3b):  they   
 zoom into the inner region around the YSO until 
the radius at which the dust becomes optically thick.
Moreover, the size of $\sim0\farcs2$ is based on the $uv$ fit  of three  low-energy lines. Higher energy lines, like those that drive the LTE  fit of CH$_3$OH here, likely trace a more compact region. We note that a smaller source size would result in a similar temperature  and a larger column density in the LTE fit since lines are optically thin.

\subsection{CH$_3$OH kinematics}\label{kin}

To study the kinematics of the hot gas traced by CH$_3$OH,
we performed fits of the visibilities for various velocity channels  for the
$7_1 - 6_1$-$A$,\,$\varv_t=0$ and 
  $12_{-1}-12_0$-$A$,\,$\varv_t=0$  lines
assuming that the emission follows an elliptic Gaussian distribution in space.
In this case, we excluded the $ 14_{-1} - 14_0$-$A$,\,$\varv_t=0$ transition because of blending  (Fig.\,\ref{fig2}).
The distributions of the velocity centroids  are shown in Fig.\,\ref{uvfit}. In the following, we assume a systemic velocity, $V_{\rm sys}$, of +1.7\,km\,s$^{-1}$ \citep{2014ApJ...786..114L}.
The lines have a clear velocity gradient in a direction parallel and very close to the equatorial plane. The high-velocity channels of the  $ 7_1 - 6_1$-$A$,\,$\varv_t=0$ line (and to some extent also of the $12_{-1}-12_0$-$A$,\,$\varv_t=0$ line)  move out of the equatorial plane in the direction of the red-shifted lobe of the jet. In Fig.\,\ref{fig4} we compare the centroid positions of the velocity channels of the $ 7_1 - 6_1$-$A$,\,$\varv_t=0$ line with those of C$^{17}$O(2--1).
CH$_3$OH behaves very differently from  C$^{17}$O: 
at low velocities ($\pm$ 1.5 km s$^{-1}$, from $V_{\rm sys}$) C$^{17}$O traces
rotating envelope/outflow cavities, while at higher velocities (up to $\pm$ 3\,km\,s$^{-1}$) it moves on the equatorial plane and shows a velocity pattern compatible with Keplerian rotation around a $0.2-0.3$\,M$_\odot$ YSO
\citep{2014ApJ...786..114L,2014A&A...568L...5C}. 
On the contrary, for the two CH$_3$OH lines analysed here,
the blue- and red-shifted velocity centroids are shifted roughly symmetrically on either side of the jet axis, indicating that the line-of-sight velocity beyond $\sim0.7$\,km\,s$^{-1}$ from systemic is dominated by rotational motions. The fact that this rotation velocity increases moving away from the protostar further indicates that  CH$_3$OH is not associated with a Keplerian disk or rotating-infalling cavity, and it is more likely associated  with an outflow/jet system. This is supported by the elongation of the integrated  emission  with a PA close to the jet axis (Table\,\ref{tab_uvfit}), and by the fact that the blue-shifted centroids are clearly shifted above the disk mid plane (Fig.\,\ref{uvfit}).

\section{Discussion}\label{origin}

\begin{figure}
\centering
\includegraphics[width=8cm]{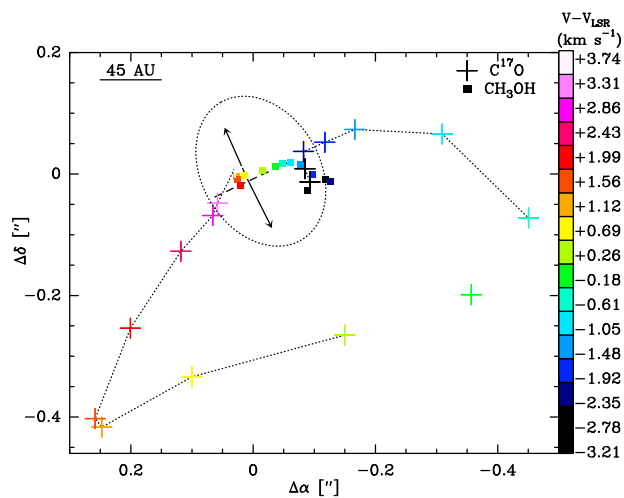}
\caption{Distribution of the centroid positions  of various velocity channels
  of the $7_1 - 6_1$-$A$,\,$\varv_t=0$ line (squares) and of the C$^{17}$O(2--1) transition (crosses). Velocities are colour-coded  according to the scale shown in the figure and are subtracted by 
the systemic velocity. The arrows indicate the direction  of the  jet; the dashed line traces the equatorial plane. For clarity, the C$^{17}$O(2--1) velocity channels are connected by dotted lines. The dotted ellipse represents the elliptical Gaussian fit to the averaged visibilities of the $7_1 - 6_1$-$A$,\,$\varv_t=0$ line.}\label{fig4}
\end{figure}

\begin{figure}
\centering
\includegraphics[width=0.6\columnwidth]{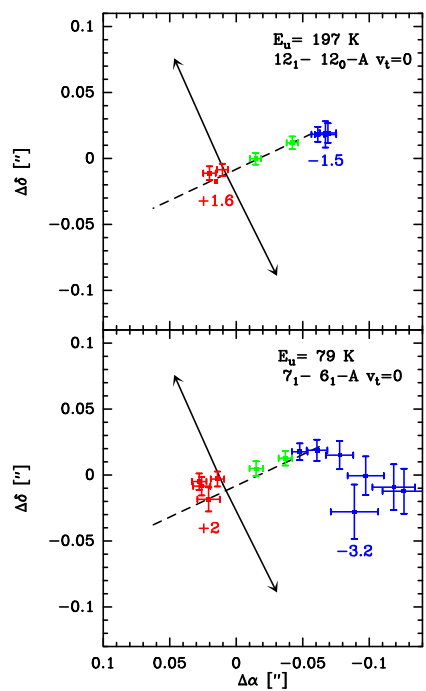}
\caption{Distribution of the centroid positions  of various velocity channels
  of the $7_1 - 6_1$-$A$,\,$\varv_t=0$ (bottom) and 
  $ 12_{-1}-12_0$-$A$,\,$\varv_t=0$ (top) lines. In both panels, the velocity of the most blue- and red-shifted channels  are subtracted by 
the systemic velocity. The velocity channels close to systemic velocity  are marked in green.
The arrows indicate the direction of the SiO jet; the dashed line traces the equatorial plane.}\label{uvfit}
\end{figure}

The CH$_3$OH emission  arises from a region of $100-150$\,mas (45--68\,AU) in radius around
 the protostar (see Table\,\ref{tab_uvfit}) and it is more compact than C$^{17}$O(2--1) (Fig.\,\ref{fig4}).
This difference seems to rule out
the association of CH$_3$OH with the cavities of the outflow traced by C$^{17}$O.  
Given the small size, CH$_3$OH could trace
the base of the low-velocity outflow seen in SO  \citep{2015A&A...581A..85P}. 
\citet{2016A&A...586L...3C} speculated that  disk winds are present in this source
based on the HDO emission,  which hinted at optically
thick emission from a  very small (18--37 AU) and dense ($n\ge10^9$\,cm$^{-3}$) slow outflowing gas. Our analysis strengthens this scenario: CH$_3$OH originates from  compact  (solid upper limit of $135$\,AU to the diameter of the emitting region), hot
($T\sim 295$\,K) gas elongated along the direction of the  jet.
We speculate that the observed velocities of CH$_3$OH are higher towards the outer part of the region than closer to the rotation axis (as expected in a disk wind model) because of beam dilution effects as described by \citet{2004A&A...416L...9P} for [OI]$\lambda$6300 data of the classical T Tauri star DG Tau. The fact that the velocity seems to increase at increasing distance from the source suggests that CH$_3$OH traces ejected gas rather than swept-up material. 
Interesting observations of high-mass YSOs \citep[e.g.,][]{2015A&A...583L...3S}  show that IR-pumped Class\,II CH$_3$OH masers  trace  the outer launching region of the primary outflow. Higher angular resolution is necessary to investigate whether the SiO and the CH$_3$OH emission seen in HH\,212 trace different velocity components of a nested onion-like system or    two different physical structures.
If CH$_3$OH is associated with an axisymmetric, steady, magneto-centrifugally accelerated disk wind, we can estimate its launching radius from Eq.\,5 of \citet{2003ApJ...590L.107A}. Assuming a poloidal velocity\footnote{For a maximum velocity CH$_3$OH of $+2$\,km\,s$^{-1}$ and an inclination angle of $\sim4^\circ$ to the plane of the sky \citep{1998ApJ...507L..79C}} of 30\,km\,s$^{-1}$ and a toroidal velocity of $1$\,km\,s$^{-1}$ (Fig.\,\ref{fig4}) at some tens of AU from the axis, the launching radius is $\sim1$\,AU, consistent with  disk wind models of water   in Class 0 YSOs \citep{2016A&A...585A..74Y}.

\section{Conclusions}
A simple cartoon (not to scale) of  the inner region of  HH\,212 is given in Fig.\,\ref{figa3}.  The bulk of the C$^{17}$O emission traces the protostellar envelope ($\sim 460$\,AU) flattened in the equatorial plane.  Low-velocity C$^{17}$O   is associated with rotating  cavity walls carved by the large scale outflow. The primary jet is shown by the SiO emission  \citep[and also detected at high-velocity in CO(3--2) and SO,][]{2015ApJ...805..186L,2015A&A...581A..85P}.  We speculate that CH$_3$OH and HDO come from a compact  ($<135$\,AU in diameter) warm ($T\sim 295$\,K) region likely associated with a disk wind gas accelerated
at the base. Indeed, HDO and CH$_3$OH have similar peak velocities ($\sim 2$\,km\,s$^{-1}$) and line-widths (4--5\,km\,s$^{-1}$). If CH$_3$OH traces a steady, axisymmetric, magneto-centrifugally driven disk wind, the launching region is at a radius of  $\sim1$\,AU from the YSO.

\begin{acknowledgements}
  The authors  thank the referee for the comments that have helped improve the clarity of the paper.  S.L. acknowledges fruitful discussions with A. Sanna and F. Fontani. This paper makes use of the following ALMA data: ADS/JAO.ALMA\#2011.0.000647.S. ALMA is a partnership of ESO (representing its member states), NSF (USA), and NINS (Japan), together with NRC (Canada), NSC and ASIAA (Taiwan), and KASI (Republic of Korea), in cooperation with the Republic of Chile. The Joint ALMA Observatory is operated by ESO, AUI/NRAO, and NAOJ.
\end{acknowledgements}
\bibliographystyle{aa}

\begin{appendix}
  \section{Additional material}
Table\,\ref{tab_id} lists all methanol emission lines observed towards HH\,212--MM1.

Table\,\ref{tab_lte} presents the best fit results and the 95\% highest probability density (HPD) interval ranges. 

\begin{figure*}[]
\centering
\subfigure{\includegraphics[width=13cm, bb =0 0 769 497,clip]{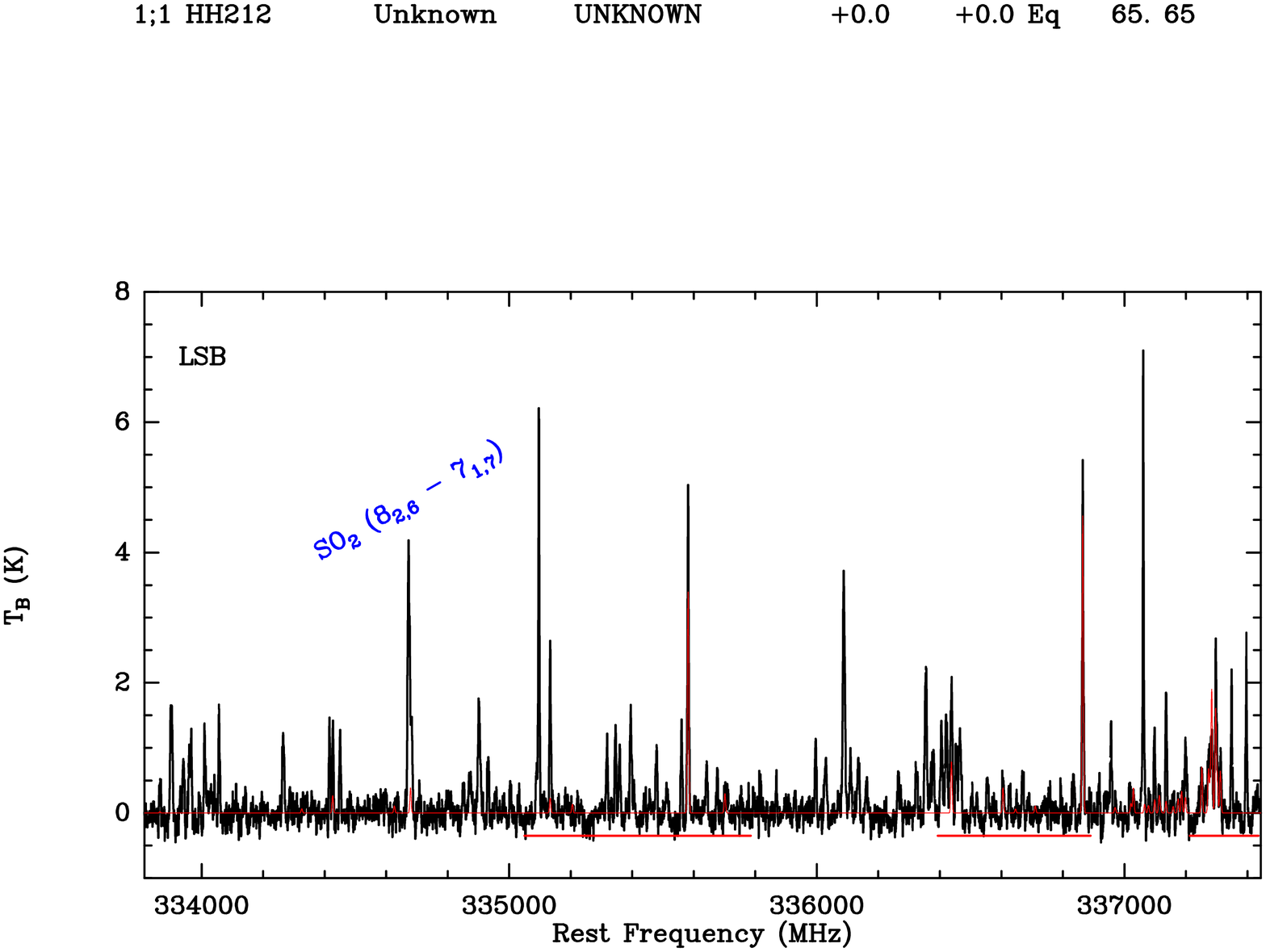}}
\subfigure{\includegraphics[width=13cm, bb =0 0 769 497,clip]{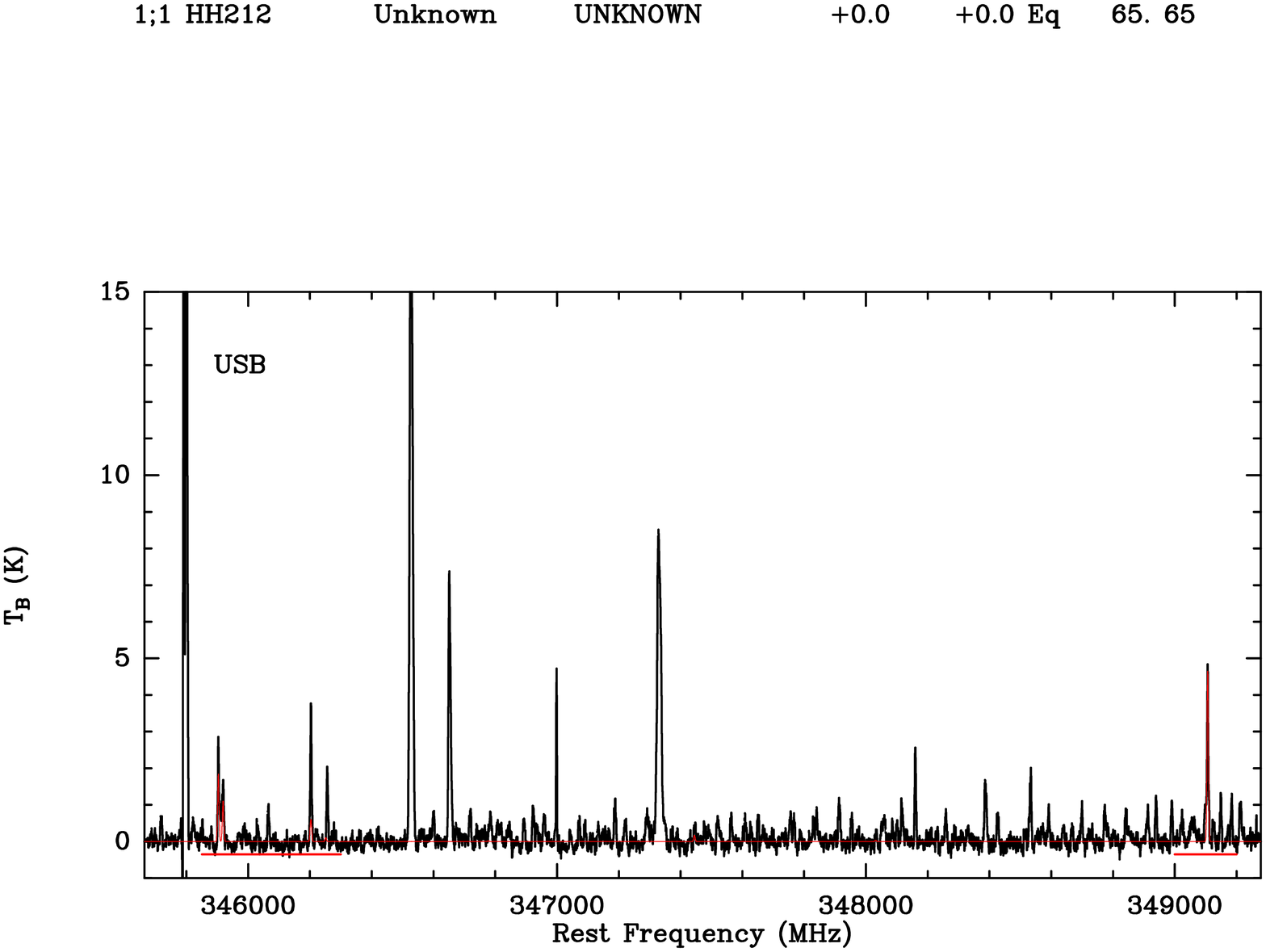}}
\caption{Full spectrum extracted at the dust peak position HH\,212-MM1 (top panel: lower side band data; lower panel: upper side band data). The horizontal red lines mark the regions of the spectrum plotted in Fig.\,\ref{fig2} where most of the methanol lines are found. The best fit LTE synthetic spectrum of methanol is displayed in red over the full 8\,GHz bandwidth.    In the upper panel, the SO$_2 (8_{2,6}- 7_{1,7})$ is marked because it overlaps with a CH$_3$OH line (the $25_{-3}-24_{-2}-E, \varv_t=1$ transition, $E_{\rm u}$= 1078\,K). This methanol line was not included in the fit.}\label{figa1}
\end{figure*}

  Figure\,\ref{figa1} shows the full spectrum (both upper side band (USB), top panel, and lower side band (LSB), lower panel) extracted at the position 
of the HH\,212-MM1 protostar. The horizontal red lines mark the regions of the spectrum plotted 
in Fig.\,\ref{fig2} where the majority of methanol lines are found. 

\begin{figure}[]
\centering
\includegraphics[width=9cm]{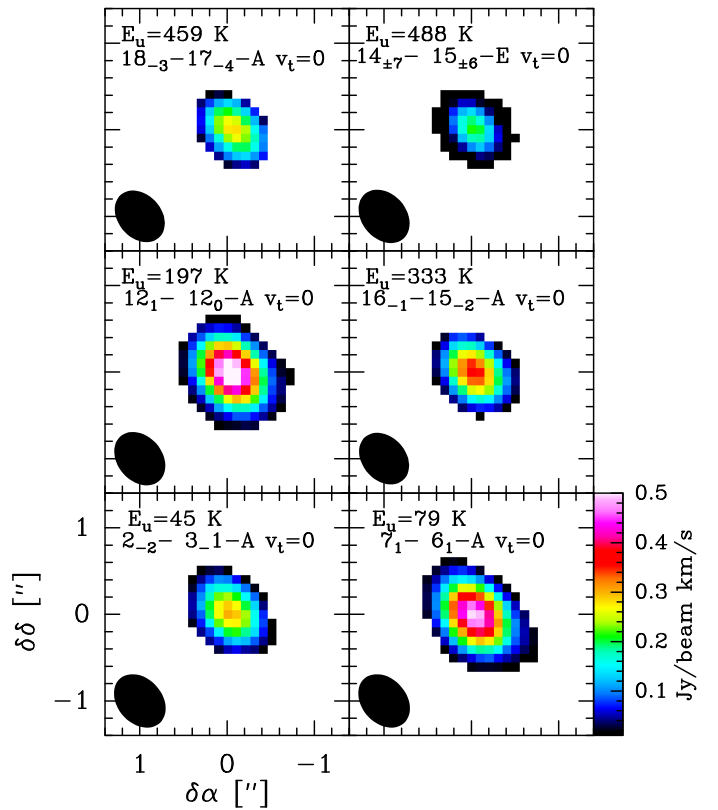}
\caption{Integrated intensity maps of different methanol transitions. In each panel, the filled ellipse shows the synthesised beam of the corresponding map. To identify the lines, the upper energy of each transition is also indicated.}\label{fig7}
\end{figure}

\begin{figure}
\centering
\includegraphics[bb= 14 46 576 794,clip,width=0.37\textwidth,angle=90]{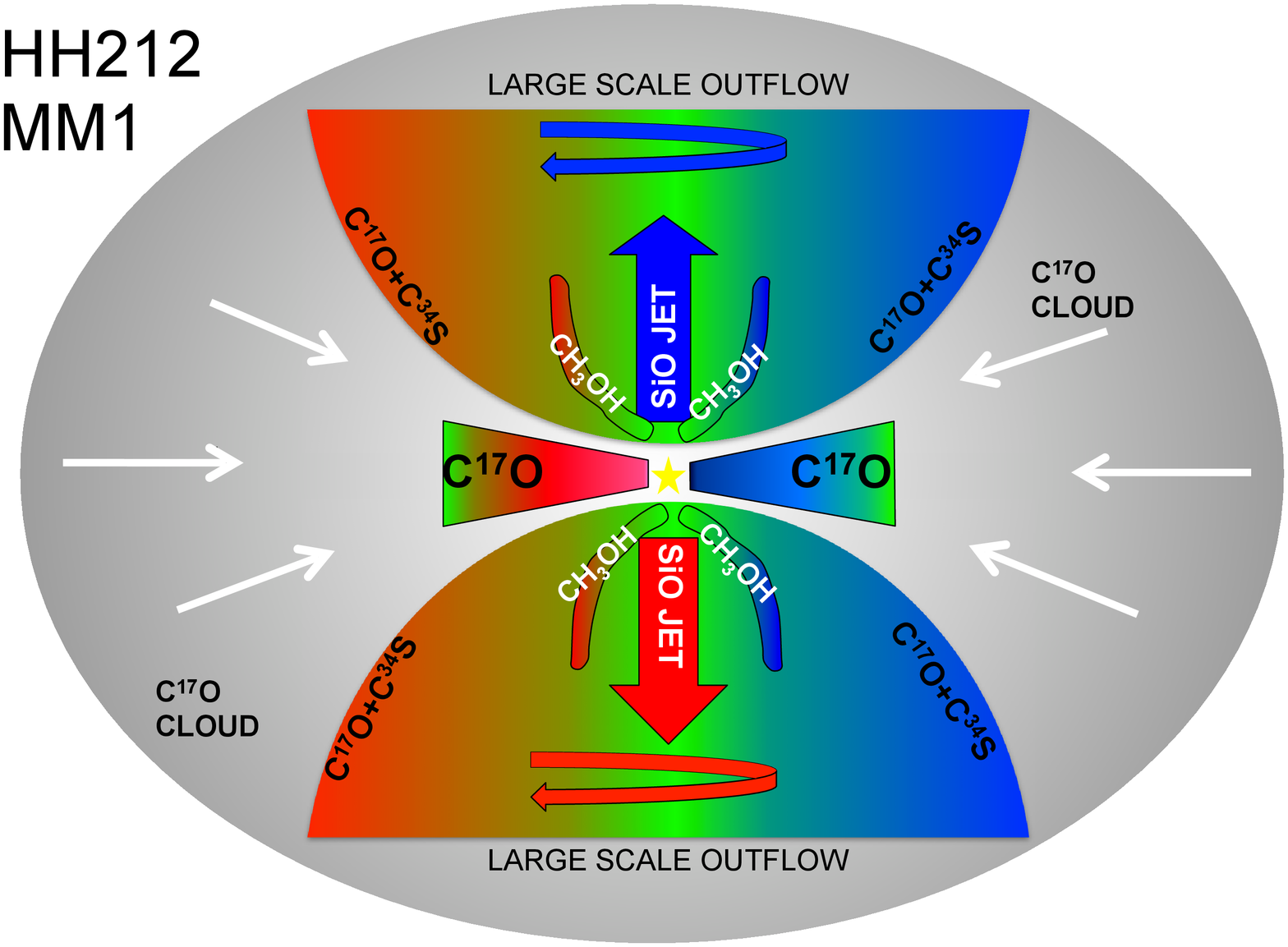}
\caption{Cartoon (not to scale) illustrating the scenario proposed for the inner region of the HH\,212 protostellar system based on the results presented in this paper and by \citet{2014A&A...568L...5C}.}\label{figa3}
\end{figure}

Figure\,\ref{fig7} shows the maps of the  integrated line intensity of different methanol lines not affected by blending with other transitions.  
Figure\,\ref{figa3} summarises the scenario proposed for the inner region of the HH\,212 protostellar system (not to scale).

 \begin{table}
\caption{List of CH$_3$OH transitions identified towards the position of the
HH\,212-mm protostar.}   
\label{tab_id}
\centering
\begin{tabular}{l c r c}  
\hline\hline                 
Transition & $\nu\tablefootmark{a}$ & $E_{\rm up}\tablefootmark{a}$ & $A_{i,j}\tablefootmark{a}$\\
&\multicolumn{1}{c}{(MHz)}&\multicolumn{1}{c}{(K)}&\multicolumn{1}{c}{($10^{-4}\,\rm{s}^{-1}$)}\\
\hline                       
     $  2_{-2} - 3_{-1}$-$A$,\,$\varv_t=0$&   335133.570 &   44.7& 0.27 \\       
     $  7_1    - 6_1$-$A$,\,$\varv_t=0$  &   335582.017 &    79.0&1.63  \\       
     $ 14_\pm7    - 15_\pm6$-$A$,\,$\varv_t=0$ &   336438.224 &   488.2&0.36  \\       
     $  7_1    - 6_1$-$A$,\,$\varv_t=2$  &   336605.889 &   747.4&1.64  \\       
     $ 12_{-1}  - 12_0$-$A$,\,$\varv_t=0$&   336865.149 &   197.1&2.04  \\           
     $  7_{-3} - 6_{-3}$-$A$,\,$\varv_t=2$&   337252.172 &   722.8&1.39  \\       
     $  7_3    - 6_3 $-$A$,\,$\varv_t=2$ &   337252.173 &   722.8&1.39  \\       
     $  7_4    - 6_4 $-$A$,\,$\varv_t=2$ &   337273.561 &   679.3&1.13   \\       
     $  7_{-2}  - 6_{-2}$-$E$,\,$\varv_t=2$&  337279.180 &   709.7&1.54  \\       
     $  7_0    - 6_0 $-$A$,\,$\varv_t=2$ &   337284.320 &   573.0&2.29  \\       
     $  7_3    - 6_3 $-$E$,\,$\varv_t=2$ &   337295.913 &   686.2&1.37  \\       
     $  7_1    - 6_1$-$A$,\,$\varv_t=1$ &    337297.484 &   390.0&1.66  \\       
     $  7_2    - 6_2$-$E$,\,$\varv_t=2$ &    337302.644 &   651.0&1.55  \\       
     $  7_{-1} - 6_{-1}$-$E$,\,$\varv_t=2$&   337312.360 &   596.8&1.65  \\       
     $ 16_{-1} - 15_{-2}$-$A$,\,$\varv_t=0$&  345903.916 &   332.7&0.90  \\       
     $ 18_{-3} - 17_{-4}$-$E$,\,$\varv_t=0$&  345919.260 &   459.4&0.70  \\       
     $  5_{-4} - 6_{-3}$-$A$,\,$\varv_t=0$ &  346202.719 &   115.2&0.22  \\       
     $  5_4    - 6_3$-$A$,\,$\varv_t=0$   &  346204.271 &   115.2&0.22  \\       
     $ 14_{-1} - 14_0$-$A$,\,$\varv_t=0$  &  349106.997  &   260.2&2.20  \\       
\hline 

\end{tabular}
\tablefoot{
  \tablefoottext{a}{From the JPL database \citep{pickett_JMolSpectrosc_60_883_1998}}
}
 \end{table}
 
\begin{table}
  \caption{Methanol LTE fit results}\label{tab_lte}
\begin{tabular}{l l c}  
\hline\hline                 
&Best Fit & 95\% HPD\\
Temperature (K)& 295&290--300\\
Column density ($10^{17}$\,cm$^{-2}$)& 3.1&3.0--3.2\\
Source size (\arcsec)&0.2&---\\
Line-width  (km\,s$^{-1}$)&4.6&4.5--4.7\\
Peak velocity  (km\,s$^{-1}$)&2.1&2.0--2.1\\
\hline                       
\end{tabular}

\end{table}

\end{appendix}
\end{document}